\documentclass[12pt]{article}
\usepackage{graphicx}
\usepackage{amsmath}
\usepackage{amssymb}
\usepackage{bbm}

\renewcommand{\Re}{\operatorname{Re}}

\newcommand{\Tr}{\operatorname{Tr}}
\newcommand{\sgn}{\operatorname{sgn}}
\newcommand{\Centre}{\operatorname{Centre}}
\newcommand{\R}{{\mathbbm{R}}}
\newcommand{\Z}{{\mathbbm{Z}}}
\newcommand{\id}{{\mathbbm1}}
\newcommand{\dd}{{\mathrm{d}}}

\newcommand{\ii}{{\mathrm{i}}}
\newcommand{\ee}{{\mathrm{e}}}
\newcommand{\F}{{\mathrm{F}}}
\newcommand{\A}{{\mathrm{A}}}
\newcommand{\V}{{\mathrm{V}}}

\newcommand{\Wilson}{{\text{Wilson}}}
\newcommand{\Villain}{{\text{Villain}}}

\newcommand{\U}{{\mathrm{U}}}
\newcommand{\SU}{{\mathrm{SU}}}
\newcommand{\SO}{{\mathrm{SO}}}
\newcommand{\SUNZN}{\SU(N)/\Z_N}
\newcommand{\PF}{\Tr_\F P}
\newcommand{\PA}{\Tr_\A P}

\begin{document}

\begin{titlepage}

\begin{center}

\begin{flushright}
  CERN-TH/2002-312\\
  hep-lat/0211004\\[12ex] 
\end{flushright}

\textbf{\large
Comparison of SO(3) and SU(2) Lattice Gauge Theory} 
\\[6ex]

{Philippe de Forcrand$^{a,b,}$\footnote{forcrand@phys.ethz.ch}
  and Oliver Jahn$^{a,}$\footnote{jahn@phys.ethz.ch}}
\\[6ex]
{${}^a${\it Institute for Theoretical Physics, ETH Z\"{u}rich,
CH-8093 Z\"{u}rich, Switzerland}\\[1ex]
${}^b${\it CERN, Theory Division, CH-1211 Gen\`{e}ve 23, Switzerland}}
\\[10ex]
{\small \bf Abstract}\\[2ex]
\begin{minipage}{14cm}{\small
    The Villain form of SO(3) lattice gauge theory is studied and
    compared to Wilson's SU(2) theory.  The topological invariants in
    SO(3) which correspond to twisted boundary conditions in SU(2) are
    discussed and lattice observables are introduced for them.  An
    apparent SO(3) phase with negative adjoint Polyakov loop is
    explained in terms of these observables.  The electric twist free
    energy, an order parameter for the confinement-deconfinement
    transition, is measured in both theories to calibrate the
    temperature.  The results indicate that lattices with about $700^4$
    sites or larger will be needed to study the SO(3) confined phase.
    Alternative actions are discussed and an analytic path connecting
    SO(3) and SU(2) lattice gauge theory at weak coupling is exhibited.
    The relevance for confinement of the centre of the gauge group is
    discussed.}
\end{minipage}
\end{center}
\vspace{1cm}

\end{titlepage}

\section{Motivation}

Although SO(3) lattice gauge theories (LGT) are expected to have the
same continuum limit as theories with gauge group SU(2), they are
interesting for at least two reasons.  First, the equivalence of the
non-perturbative continuum limits has not been proven yet; there are in
fact some indications that they might be different.  Second, they
provide a tool to study the role of the centre of the gauge group,
which is often considered to be closely related to confinement: the
centre of SO(3), as opposed to that of SU(2), is trivial.

The expectations on the continuum limit are largely based on the
\emph{naive} continuum limit as obtained by expanding the classical
action around its minima.  As $\SO(3)\cong\SU(2)/\Z_2$ is locally
indistinguishable from SU(2), in particular in any small region around
the identity, plaquette actions based on SO(3) and SU(2) link variables
have the same naive continuum limit.  Invoking standard arguments of
universality of the continuum limit, one expects the non-perturbative
continuum limit to be the same, too.  However, this expectation is not
based on a rigorous theorem and should be checked for each individual
case.  Studies of the Stefan--Boltzmann law in both theories for
instance have raised doubts \cite{LR}.

Studies of mixed actions containing both SU(2) and SO(3) parts [the
latter being considered as the adjoint representation of SU(2)] have
not yet been able to demonstrate this universality.  Rather they reveal
a line of first order phase transitions separating the weak coupling
regions of the pure SO(3) and SU(2) theories \cite{BhC,CHS}, so the
continuum limits are not analytically connected in this plane of
couplings.  The location of the phase transitions is essentially
independent of the lattice size, so they are considered as bulk.  In
the pure SO(3) theory, the transition occurs at very weak coupling.
This puts an upper bound on the lattice spacing in the weak coupling
phase, and prohibitively large lattices are required to study low
temperatures.

A further peculiarity of the SO(3) theory is the occurance of a new
meta-stable ``phase'' (long-lived state in the Monte-Carlo simulation)
with a negative expectation value of the adjoint Polyakov loop at weak
coupling \cite{CS,DG1}.  Apart from the Polyakov loop, this ``phase''
seems to be very similar to the standard phase with positive adjoint
Polyakov loop.  Still its existence has fuelled speculations about
differences between SO(3) and SU(2) theories.  We will show that the
mysterious phase can be understood in terms of the SU(2) theory.

The question of whether SO(3) and SU(2) LGT describe the same quantum
theory is also important for our understanding of the role of the
centre of the gauge group.  A common belief asserts that deconfinement
is caused by the breakdown of the centre symmetry \cite{SY}.  However,
it has been argued that this symmetry is a peculiarity of the lattice
theory and not present in other regularisations \cite{Smilga}.  In the
SO(3) lattice theory, the centre symmetry is absent and cannot be
broken.  If this theory has a confinement-deconfinement transition,
too, it has to be explained by a different mechanism.  

While the centre symmetry is absent, it should be noted that centre
vortices do exist in an SO(3) theory: their definition actually does not
rely on the centre but on a non-trivial first homotopy group of the
gauge group modulo its centre.  As $SO(3)$ is $SU(2)$ modulo its
centre $\Z_2$, this is the same for both groups,
$\pi_1[\SU(2)/\Z_2]=\pi_1[\SO(3)]=\Z_2$.  So confinement mechanisms
based on centre vortices can apply to both theories equally well.

In the absence of the centre symmetry, the question of an order
parameter for the confinement-deconfinement mechanism arises.  The
Polyakov loop in the fundamental representation is useless because its
sign is not determined.  The adjoint Polyakov loop is not good either,
since it has a non-zero expectation value also in the confined phase
because adjoint charges can be screened by gluons.
A useful alternative is known for the SU(2) theory.  The expectation
value of a large spatial 't Hooft loop follows a perimeter law in the
confined phase and an area law in the deconfined phase, the coefficient
defining the ``dual string tension'' \cite{tH1,dFDEP}.  A particularly
convenient way to measure 't Hooft loops is provided by twisted
boundary conditions on a hypercubic lattice (a torus).  These introduce
chromo-electric and -magnetic fluxes through the torus and correspond
to 't Hooft loops of maximal size \cite{tH2}.  Their expectation value
has been measured both at zero \cite{KT-tHooft-loop} and at finite
temperatures \cite{dFvS,vSdF}, and perimeter respectively area law have
been demonstrated.  We will show that the free energy of these fluxes
can also be measured in an SO(3) LGT.  As 't Hooft loops can be thought
of as generating centre vortices, this shows that the latter do indeed
play a role.  Preliminary accounts of the present work have been given
in \cite{StaraLesna,LAT02}.

We begin in Sec.~\ref{sec:topology} with a review of the topological
invariants of $\SU(N)$ and $\SUNZN$ lattice gauge theory on a 4-torus,
and of the implementation of twisted boundary conditions in $\SU(N)$ LGT.
Section \ref{sec:SO3-Z2} recalls a formulation of the $\SU(2)$ theory
as a coupled $\SO(3)$-$\Z_2$ theory.  This formulation motivates a
definition of twist in the $\SO(3)$ theory which is presented in
Sec.~\ref{sec:twist-SO3}.  In Sec.~\ref{sec:mystery}, an interpretation
of the ``phase'' with negative adjoint Polyakov loop is presented.
Section \ref{sec:measurement} describes our numerical computation of
the free energy of electric twist in SO(3).  Before the conclusions, a
discussion of alternative actions and an analytic path between SO(3)
and SU(2) at weak coupling are proposed.  The general relation between
Polyakov loops and twist is reviewed in the appendix.

\section{Topology in $\SUNZN$ and $\SU(N)$ on the torus}
\label{sec:topology}

In the continuum, there is no local difference between $\SU(N)$ and
$\SUNZN$ gauge fields.  The gauge potential can be expressed either in
the fundamental representation (with generators $t^a$) or in the
adjoint one (with generators $T^a$),
\begin{equation}
  \label{eq:A}
  A^{\F}_\mu = A^a_\mu t^a 
  \quad\text{or}\quad
  A^{\A}_\mu = A^a_\mu T^a \;.
\end{equation}
The components $A^a_\mu$ are always the same.  The only difference is
in the boundary conditions, since the transition functions
$\Omega_\mu$ take values in the gauge group.  On a torus with extent
$L_\nu$ in direction $e_\nu$:
\begin{equation}
  \label{eq:bc}
  A^{\F/\A}_\mu(x+L_\nu e_\nu) = 
  \Omega^{\F/\A}_\nu(x) \,
  \bigl( A^{\F/\A}_\mu(x) - \ii \partial_\mu \bigr) \,
  \Omega_\nu^{\F/\A\,\dagger}(x) \;.
\end{equation}
The transition functions satisfy the consistency conditions
\begin{align}
  \label{eq:ccF}
  \Omega^{\F}_\mu(x+L_\nu e_\nu) \, \Omega^{\F}_\nu(x) &= 
  z_{\mu\nu} \, \Omega^{\F}_\nu(x+L_\mu e_\mu) \, \Omega^{\F}_\mu(x) \\
  \label{eq:ccA}
  \Omega^{\A}_\mu(x+L_\nu e_\nu) \, \Omega^{\A}_\nu(x) &= 
  \Omega^{\A}_\nu(x+L_\mu e_\mu) \, \Omega^{\A}_\mu(x) \;.
\end{align}
The twist $z_{\mu\nu}=\exp\bigl(\frac{2\pi\ii}{N}n_{\mu\nu}\bigr)$ with
integer $n_{\mu\nu}$ has to be an element of the centre $\Z_N$ of
$\SU(N)$.  The group $\SUNZN$ has trivial centre, so no twist is
allowed for $\Omega^{\A}_\mu$.  With (\ref{eq:ccF}) and (\ref{eq:ccA}),
both descriptions are equivalent also globally, as any set of
transition functions in $\SUNZN$ can be lifted to $\SU(N)$ and any in
$\SU(N)$ can be projected to $\SUNZN$.  The twist $z=\{z_{\mu\nu}\}$,
specified by the consistency conditions in the case of $\SU(N)$ fields,
becomes an ordinary topological invariant like the instanton number for
$\SUNZN$ fields.  In a slight abuse of language, we shall call it twist
also there.  In addition to the 6 twists
$z_{\mu\nu}\in\pi_1[\SUNZN]=\Z_N$ which are associated with
two-dimensional sections of the torus ($\mu\nu$-planes), there is one
more topological invariant $q\in\pi_3[\SUNZN]=\Z$ associated with the
bulk \cite{Sedlacek,Nash}.  The usual topological charge (instanton
number) $Q$ is a combination of both,
\begin{equation}
  Q = q - \frac{\kappa}{N}
  \quad\text{with}\quad
  \kappa =
  \tfrac{1}{8}\varepsilon_{\mu\nu\rho\sigma}n_{\mu\nu}n_{\rho\sigma}
  \;.
  \label{eq:Q}
\end{equation}
It can take fractional values for non-zero $n$.

On a lattice, periodic link variables represent all possible boundary
conditions with trivial consistency conditions [i.e.~(\ref{eq:ccA}) or
(\ref{eq:ccF}) with $z_{\mu\nu}=1$].  Therefore, the $\SUNZN$ lattice
theory (with periodic boundary conditions) automatically contains all
``twist'' and instanton sectors while the periodic $\SU(N)$ theory
contains all instanton sectors but only trivial twist.  A non-trivial
(fixed) twist $z$ can be implemented by modifying the periodic boundary
conditions to $U_\nu(x+L_\mu)=z_{\mu\nu} U_\nu(x)$ for $\mu<\nu$ for a
single value of $x_\mu$.  Alternatively, one can work with periodic
link variables by absorbing the factor $z_{\mu\nu}$ into one plaquette
in each $\mu\nu$-plane \cite{GJKA}.  The partition function of $\SU(N)$
LGT with Wilson action for twist $z$ then becomes
\begin{equation}
  \label{eq:Z-tbc}
  Z^\Wilson_{\SU(N)}(z) = \int_{\SU(N)} \prod_l\dd U_l \exp \biggl[
  - \beta \sum_p \Bigl( 1 - \frac{1}{N} \Re (\zeta_p \Tr_\F U_p)
  \Bigr) \biggr]
\end{equation}
where $l$ labels links, $p$ plaquettes and
\begin{equation}
  \label{eq:alpha-z}
  \zeta_{x,\mu\nu} = 
  \begin{cases}
    z_{\mu\nu} & \text{if } x_\mu=x_\nu=0 \;, \\
    1 & \text{otherwise.}
  \end{cases}
\end{equation}

In the $\SUNZN$ lattice theory, the twist cannot be imposed in this way
by the boundary conditions -- all twists $z$ are included within
periodic boundary conditions.  A ``field-theoretic'' definition is not
possible
either, since $z$ does not have a local representation as an integral
like the topological charge ($Q\propto\int F\tilde F$).%
\footnote{%
  In terms of $\U(N)$ fields they do \cite{LPR}, but we will not pursue
  this any further.}
However, an $\SUNZN$ lattice theory can also be described redundantly
in terms of $\SU(N)$ variables.  Just imagine mapping each link variable
to an $\SU(N)$ element, with arbitrary $\Z_N$ phase.  The resulting
link configuration will in general not be periodic, but only periodic
modulo $\Z_N$.  The transition functions will satisfy twisted
consistency conditions like (\ref{eq:ccF}), but the twist will be a
priori arbitrary.  One way of assigning a unique value, would be some
smoothness requirement.  One could, for instance, try to fix the
ambiguity by demanding that the $\SU(N)$ plaquette variables be as close
to $\id$ as possible.  This is not natural, however, since $\SUNZN$
lattice gauge fields describe continuum gauge fields with $\Z_N$
monopoles.  Attempting a ``smooth'' lift to $\SU(N)$ ignores the
latter.  We therefore choose a different approach, defining twist by
analogy with the $\SU(N)$ theory in a formulation in $\SUNZN$ and $\Z_N$
variables.

\section{$\SU(2)$ LGT as a coupled $\SO(3)$-$\Z_2$ theory}
\label{sec:SO3-Z2}

In order to carry the definition of twist over to an $\SUNZN$ LGT, we
invoke a formulation of the $\SU(N)$ theory in terms of $\SUNZN$ and
$\Z_N$ variables.  We specialise to $N=2$, the case studied numerically
in this paper.

The Wilson partition function for twist $z$ (\ref{eq:Z-tbc}) can be
expressed as \cite{MP,Tomboulis,KT-weak,AH}
\begin{equation}
  \label{eq:SO3-Z2}
  Z^\Wilson_{\SU(2)}(z) = \mathcal{N} \sum_{\alpha_p=\pm1} \int \prod_l\dd U_l 
  \exp \biggl( \frac{\beta}{2} \sum_p \alpha_p \Tr_\F U_p \biggr) \,
  \prod_c \delta(\alpha_c,1) \,
  \prod_{\mu<\nu} \delta(\alpha_{\mu\nu},z_{\mu\nu}) \;.
\end{equation}
where $c$ labels cubes and
\begin{align}
  \label{eq:alpha-c}
  \alpha_c &= \prod_{p\in\partial c} \alpha_p \;, \\
  \label{eq:alpha-munu}
  \alpha_{\mu\nu} &= \prod_{x_\mu,x_\nu} \alpha_{x,\mu\nu}
\end{align}
with $x_\rho$ fixed for $\rho\ne\mu,\nu$ in the last equation.  The
factor $\mathcal{N}$ represents a change of normalisation.  The choice of
$\Z_2$ plaquette variables here is different from that of
\cite{Tomboulis,KT-weak}: they are related as $\alpha_p=\sigma_p\sgn\Tr
U_p$.

Equation (\ref{eq:SO3-Z2}) can be understood as follows.  The first set
of constraints (on $\alpha_c$) can be interpreted as the Bianchi
identity for the field strength $\alpha_p$ of a $\Z_2$ gauge theory.
It forbids monopoles and ensures local integrability: $\alpha_p$ can be
obtained from a $\Z_2$ link field (potential) $\gamma_l$,
\begin{equation}
  \label{eq:Z2-potential}
  \alpha_p = \prod_{l\in\partial p} \gamma_l
  \rlap{\qquad(locally) .}
\end{equation}
This is possible whenever the product of $\alpha_p$ over all closed
surfaces is 1.  In $\R^4$, all closed surfaces can be constructed from
elementary cubes, so the first set of constraints suffices.  On a
torus, there are 6 independent winding surfaces which cannot be
obtained in this way.  The second set of constraints fixes the products
of $\alpha_p$ on these.  If $z_{\mu\nu}=1$, $\alpha_p$ can then be
integrated globally.  Otherwise, one has to make $\alpha_{\mu\nu}$
trivial first.  This is achieved by dividing $\alpha_p$ by any
representative with the same $\alpha_{\mu\nu}$, for instance
(\ref{eq:alpha-z}); so the potential is introduced as
\begin{equation}
  \label{eq:Z2-potential-tbc}
  \alpha_p = \zeta_p \prod_{l\in\partial p} \gamma_l \;.
\end{equation}
With this representation, the sum over $\alpha_p$ in (\ref{eq:SO3-Z2})
can be replaced by one over $\gamma_l$.  There is no Jacobian because
the $\Z_2$ gauge orbits all have the same size.  The constraints are
now redundant, so $\gamma_l$ only appears in the action.  This
dependence can be removed by a change of variables
$U_l\to\gamma_l^{-1}U_l$ which replaces $\alpha_p$ by the fixed
$\zeta_p$.  Due to the invariance of the Haar measure, the $\gamma_l$
sum can then be done trivially, and the Wilson partition function
(\ref{eq:Z-tbc}) is recovered.

\section{Twist in $\SO(3)$ LGT}
\label{sec:twist-SO3}

The bridge to $\SO(3)$ is now built by noting that, without the
constraints, (\ref{eq:SO3-Z2}) becomes the Villain partition function
of $\SO(3)$ LGT \cite{Yoneya},
\begin{equation}
  \label{eq:Z-Villain}
    Z^\Villain_{\SO(3)}
    = \sum_{\alpha_p=\pm1} \int_{\SU(2)} \prod_l\dd U_l 
    \exp \biggl( \frac{\beta}{2} \sum_p \alpha_p \Tr_\F U_p \biggr)
    = \int\prod_l\dd U_l \exp (-S_\V[U]) \;.
\end{equation}
The Villain action, obtained after summation over $\alpha_p$,
\begin{equation}
  \label{eq:S-V}
  S_\V[U] = -\ln \bigl[ 2\cosh\bigl(\tfrac{\beta}{2}\Tr_\F U_p\bigr) \bigr]
\end{equation}
is invariant under $U_l\to -U_l$ for each link separately, so
(\ref{eq:Z-Villain}) defines an $\SO(3)$ LGT, redundantly formulated in
terms of $\SU(2)$ variables.

A comparison of Eqs.~(\ref{eq:SO3-Z2}) and (\ref{eq:Z-Villain})
exhibits the differences between $\SU(2)$ and $\SO(3)$ LGT.  The first
set of constraints in (\ref{eq:SO3-Z2}) shows that the $\SO(3)$ theory
contains local degrees of freedom not present in $\SU(2)$: the $\Z_2$
monopoles of the field $\alpha_p$.  Loosely speaking, these monopoles
are responsible for the bulk phase transition separating phases with
and without monopole condensation in $\SO(3)$ LGT.  Since $\alpha_p$ is
only an auxiliary field, it is more appropriate to say that the
$\SO(3)$ monopoles
\begin{equation}
  \label{eq:eta-c}
  \eta_c = \prod_{p\in\partial c} \sgn\Tr_\F U_p
\end{equation}
condense at small $\beta$.  In $\SU(2)$, these are strongly suppressed
because negative plaquettes carry a large action.

The second set of constraints shows that, as opposed to $\SU(2)$, the
partition function of $\SO(3)$ LGT with periodic boundary conditions
contains all possible topological sectors, with arbitrary instanton
number and twist $z$.  This is as expected because in $\SO(3)$ all
sectors are represented by transition functions with standard
consistency conditions Eq.~(\ref{eq:ccA}).

These considerations lead us to a natural definition of the twist $z$
in $\SO(3)$ on the lattice.  In the formulation (\ref{eq:SO3-Z2}) of
$\SU(2)$, the twist is given by Eq.~(\ref{eq:alpha-munu}) in terms of
the auxiliary variables $\alpha_p$.  In SO(3), it is better to define
it in terms of~$U_p$,
\begin{equation}
  \label{eq:twist-so3}
  \eta_{\mu\nu} = \frac{1}{L_\rho L_\sigma} \sum_{x_\rho,x_\sigma}
  \; \prod_{x_\mu,x_\nu} \sgn \Tr_\F U_{x,\mu\nu}
  \qquad (\varepsilon_{\mu\nu\rho\sigma}=1)\;.
\end{equation}
The product measures the $\Z_2$ flux of the $\SO(3)$ gauge field
$U/\Z_2$ through the $\mu\nu$-plane with fixed $x_\rho$ and $x_\sigma$.
The apparent dependence on the sign of the $\SU(2)$ links $U_l$ drops
out because each link appears in two plaquettes in the product.  Thus
$\eta_{\mu\nu}$ is a proper $\SO(3)$ observable.  Contrary to
(\ref{eq:alpha-munu}), it can be defined in any $\SO(3)$ lattice gauge
theory in formulations with or without auxiliary fields $\alpha_p$.
The average over all parallel planes with a given orientation has been
introduced because in $\SO(3)$, as opposed to $\SU(2)$, the flux can
change from plane to plane due to the monopoles (\ref{eq:eta-c}).  This
means that $\eta_{\mu\nu}$ can take fractional values $-1\le
\eta_{\mu\nu}\le 1$.  A $\Z_2$-valued twist can be defined by dividing
the range of $\eta_{\mu\nu}$ into two pieces:
\begin{equation}
  \label{eq:z-eta}
  z_{\mu\nu}(\eta) = \sgn(\eta_{\mu\nu} - c_{\mu\nu})
\end{equation}
with constant $c_{\mu\nu}$.  Note that there is no symmetry
$\eta_{\mu\nu} \leftrightarrow -\eta_{\mu\nu}$, so $c_{\mu\nu}=0$ is
not singled out.  The choice of $c_{\mu\nu}$ will be discussed in more
detail later.  The ambiguity in assigning a twist $z$ to a given
configuration is similar to that of the usual topological charge in
$\SU(2)$.  It is caused by lattice artifacts (the monopoles $\eta_c$)
which make the topological sectors connected.  What is important is the
behaviour in the continuum limit $\beta\to\infty$.  Since monopoles are
exponentially suppressed in this limit, one can hope that
$z_{\mu\nu}(\eta)$ becomes independent of $c_{\mu\nu}$ there.

Consider first the naive (classical) continuum limit.  In sectors with
so-called orthogonal twist, $\kappa=0\bmod N$ [cf.~(\ref{eq:Q})], the
configuration with lowest action (``classical ground state'') actually
has zero action even though twist is present \cite{AF,GJKA}.  In
$\SU(2)$, this ``twist eater'' can be constructed as
\begin{equation}
  \label{eq:twist-eater}
  U_\mu(x) = 
  \begin{cases}
    \Gamma_\mu & \text{if } x_\mu=0, \\
    \id & \text{otherwise},
  \end{cases}
\end{equation}
where the 4 matrices $\Gamma_\mu$ satisfy
\begin{equation}
  \label{eq:t-e-conditions}
  \Gamma_\mu \Gamma_\nu \Gamma_\mu^\dagger \Gamma_\nu^\dagger = 
  z_{\mu\nu}^* \id \;.
\end{equation}
This generates the plaquette field $U_p=\zeta_p^*$ which just cancels
the factor $\zeta_p$ in the Wilson partition function with twist
(\ref{eq:Z-tbc}), so the action vanishes.  The same configuration
minimises the Villain action (\ref{eq:S-V}), because the latter is
independent of the sign of $U_p$.  The observable $\eta_{\mu\nu}$
(\ref{eq:twist-so3}), evaluated in this configuration, is identical to
the $\SU(2)$ twist $z_{\mu\nu}$ of (\ref{eq:t-e-conditions}).  The same
holds true for small fluctuations around (\ref{eq:twist-eater}), so
$\eta_{\mu\nu}$ has the proper naive continuum limit.  For
non-orthogonal twist, the lowest-action configurations are not known
analytically, but it is expected that they still satisfy $\sgn\Tr_\F
U_p=\alpha_p$ provided the lattice is fine enough.  So $\eta_{\mu\nu}$
again coincides with the $\SU(2)$ twist.

Non-perturbatively, $\eta_{\mu\nu}$ is affected by lattice artifacts
($\eta_c$) -- just as the geometric definition of the instanton number
is \cite{geom-inst}.  In our Monte-Carlo simulations, the distribution
of $\eta_{\mu\nu}$ was always divided into two well-separated peaks
close to $\pm1$ (see Fig.~\ref{fig:density1} further below), so an
unambiguous definition of twist was possible.  For very large volumes,
however, a (centre-blind) restriction of the plaquette angle might be
necessary, like for geometric definitions of topological charges.  The
$\Z_2$-invariant choice
\begin{equation}
  \label{eq:plaq-restrict}
  |\Tr_\F U_p| > \varepsilon
  \quad\text{where}\quad
  \varepsilon > 0
\end{equation}
makes the twist sectors disconnected (with $c_{\mu\nu}=0$ in
(\ref{eq:z-eta})).\footnote{%
  The discrete auxiliary variables $\alpha_p$ are of no concern here,
  as one can sum over them and work with the action (\ref{eq:S-V}).} 
For sufficiently large $\varepsilon$, lattice artifacts are completely
suppressed and $\eta_{\mu\nu}=\pm1$.  In this case, however, the
modified Villain theory becomes equivalent to an $\SU(2)$ theory with
plaquette restriction $\Tr_\F U_p>\varepsilon$, summed over all
twist sectors.  The value of this theory for a comparison of $\SO(3)$
and $\SU(2)$ lattice gauge theory is thus questionable.  We shall
\emph{not} apply a constraint like (\ref{eq:plaq-restrict}).  For the
parameters used in our simulations, the ambiguity in (\ref{eq:z-eta})
has negligible impact.

\section{Negative Polyakov loops}
\label{sec:mystery}

In this section, we would like to comment on the apparent weak coupling
phase with negative adjoint Polyakov loop observed in
Refs.~\cite{CS,DG1}.  In these articles, $\SO(3)$ lattice gauge theory
was studied with both the Wilson and the Villain action.  For the
lattice sizes used ($N_t\le 8$), only a single, first-order phase
transition was found.  This is the bulk phase transition below which
monopoles condense.  Although its position was found to be essentially
independent of $N_t$, it was also interpreted as a deconfinement
transition, because the expectation value of the adjoint Polyakov loop
is very small below the transition and has a large modulus above.
There, two different meta-stable ``phases'' were observed, depending on
the initial configuration of the Monte-Carlo simulation.  The two
phases are distinguished by the sign of the adjoint Polyakov loop.  The
positive value can be understood as a standard deconfined phase where
$\tfrac12\Tr_\F P$ has an expectation value tending to $\pm1$ in the limit
$\beta\to\infty$.  The adjoint loop $\PA=4(\tfrac12\PF)^2-1$ therefore
tends to $+3$.  No explanation was found for the other, ``mysterious''
phase.  In \cite{DG1}, the correlator of two Polyakov loops and the
response to an external field coupled to the Polyakov loop were found to
be very similar in both phases, so the authors concluded that they were
physically equivalent.  At the critical coupling, tunnellings between
both weak coupling phases and the unique strong coupling phase were
observed.

\begin{figure}[tb]
  \centering
  \mbox{\includegraphics[width=.7\textwidth,trim=-15 5 10 8]{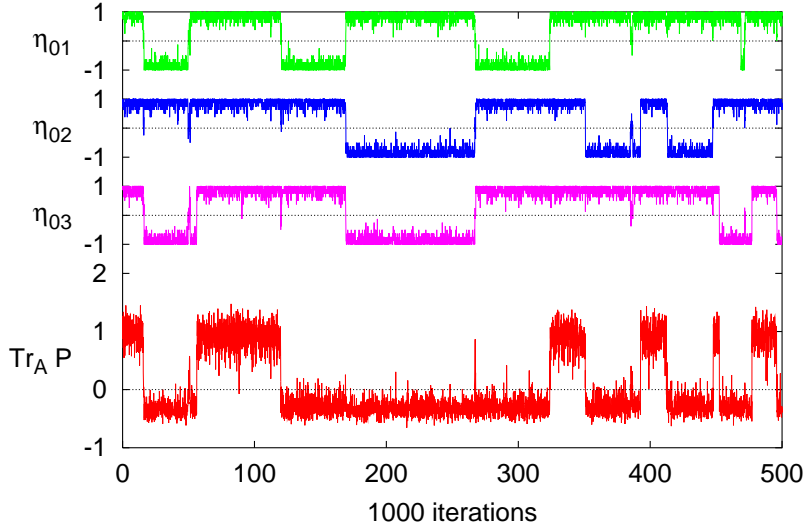}}
  \caption{Monte Carlo history of the 3 electric twist variables (top) and 
    the adjoint Polyakov loop (bottom) ($4^4$ lattice, $\beta=4.5$).
    Negative Polyakov loop values are always accompanied by one or more
    negative electric twists.}
  \label{fig:MChistory}
\end{figure}

Based on the comparison between $\SO(3)$ and $\SU(2)$ LGT, we propose
an explanation of the two weak coupling phases in terms of different
twist.  Consider the semi-classical limit.  Without twist, all values
of the Polyakov loop are compatible with zero classical action.  These
are the ``toron'' flat directions.  The effective potential localises
the fundamental Polyakov loop around two values which tend to $\pm1$ in
the continuum limit $\beta\to\infty$.  Either way, this implies a
positive adjoint Polyakov loop tending to $+3$ in the continuum limit.
With non-vanishing twist, the flat directions disappear.  For
orthogonal twist, the configuration with smallest classical action is
the twist eater (\ref{eq:twist-eater}).  The (untraced) Polyakov loop
of this configuration is just $\Gamma_0$.  The simplest solution of
(\ref{eq:t-e-conditions}) with a single electric twist $z_{03}=-1$ is
$\Gamma_0=\ii\sigma_1$ and $\Gamma_3=\ii\sigma_2$, so in this case, the
fundamental and adjoint Polyakov loops are $\Tr_\F \Gamma_0=0$ and
$\Tr_\A\Gamma_0=-1$.  In general, the twist eating matrices satisfy
$(a)$ $\Tr_\A\Gamma_0=-1$ if $z_{0i}=-1$ for some $i$, or $(b)$ $\Tr_\A
\Gamma_0=+3$ if $z_{0i}=0$ for all $i$, but $z_{i j}\ne0$ for some $i$,
$j$.  The arguments leading to this known result are collected in the
Appendix.  This classical consideration suggests that the phase with
negative Polyakov loop is one with electric twist.  This can be
verified numerically.  Figure \ref{fig:MChistory} shows the Monte-Carlo
history of the 3 electric twist variables $\eta_{0 i}$ as well as the
adjoint Polyakov loop on a $4^4$ lattice at $\beta=4.5$, just above the
bulk phase transition at $\beta\approx4.45$.
The system tunnels between phases with positive and negative Polyakov
loop.  The negative phases always have one or more electric twists and
the positive phases none.  The twist observable (\ref{eq:twist-so3})
has now been studied also with the SO(3) Wilson action, and meta-stable
phases with different $z_{0i}$ have similarly been found \cite{BBMP2}.

We would like to remark that semi-classically, the above argument is
not conclusive.  The twist eater has a Polyakov loop $\PF=0$, i.e. on the
maximum of the one-loop effective potential.  Therefore, it might not
be the dominant semi-classical configuration; a configuration with a
$\Z_N$ interface (Euclidean ``domain wall''), interpolating between the
minima of the effective potential, might have a lower effective action.
Outside of the interface, the adjoint Polyakov loop is positive.  For
large volumes, its spatial average therefore is positive, too.  If the
configuration with an interface indeed has lower effective action, the
average Polyakov loop will be positive even in the presence of electric
twist, provided the spatial extent of the lattice is sufficiently large
compared to the interface thickness.

\section{Twist free energies}
\label{sec:measurement}

We shall now measure the free energy of various twists, both in
$\SO(3)$ and in $\SU(2)$.  The results are used to calibrate the
$\SO(3)$ lattice spacing.  We choose the $\SO(3)$ coupling $\beta=4.5$
just above the bulk phase transition at $\beta\approx 4.45$ in
order to find the largest lattice spacing available in the weak
coupling phase of $\SO(3)$.

\begin{figure}[tb]
  \centering
  \mbox{\includegraphics[width=.7\textwidth]{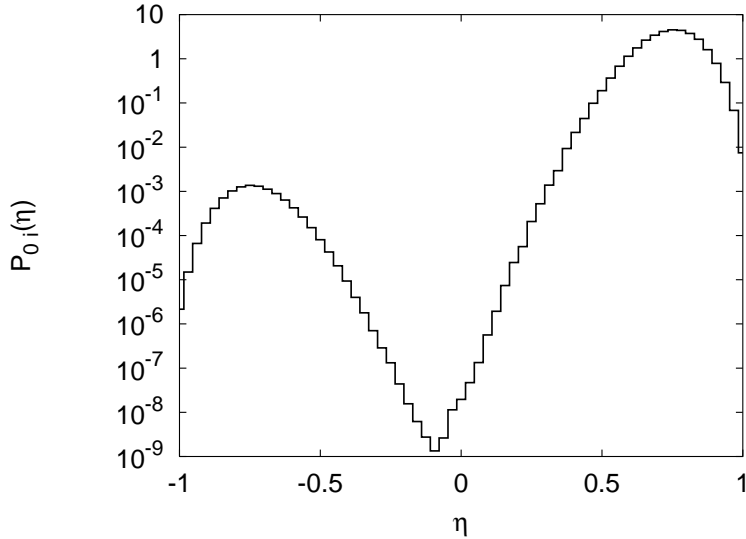}}
  \caption{``Density of states'' $P_{0 i}$ of an electric twist
    ($8^3\times 4$ lattice, $\beta=4.5$).}
  \label{fig:density1}
\end{figure}

The twist free energies $F(z)$ are defined in terms of partition functions
with and without twist,
\begin{equation}
  \label{eq:F}
  F(z) = -\ln \frac{Z(z)}{Z(1)} \;.
\end{equation}
For electric twist ($z_{0i}\ne1$), $F(z)$ is an order parameter for the
con\-fine\-ment-decon\-fine\-ment transition \cite{dFvS,vSdF}: for large
volumes, it tends to 0 in the confined phase, while diverging as
$F\sim\tilde\sigma L^2$ with the spatial extent $L$ in the deconfined
phase.  $\tilde\sigma$ is the dual string tension and depends on the
temperature.  In $\SU(2)$, the partition functions $Z(z)$ and $Z(1)$
are defined by imposing twisted and periodic boundary conditions,
respectively.  In $\SO(3)$, all twists are summed over within periodic
boundary conditions, and twist is an observable like the usual
topological charge in $\SU(2)$.  The partition functions are defined
using (\ref{eq:z-eta}),
\begin{equation}
  \label{eq:Z}
  Z(z) = Z_{\text{tot}} \biggl\langle \prod_{\mu<\nu}
  \delta(z_{\mu\nu}(\eta),z_{\mu\nu}) \biggr\rangle
\end{equation}
where $Z_{\text{tot}}=\sum_{\{z_{\mu\nu}\}}Z(z)$.  In order to
determine the value $c_{\mu\nu}$ of Eq.~(\ref{eq:z-eta}) which divides
trivial and non-trivial twist, we use the ``density of states'' as a
function of a single $\eta_{\mu\nu}$,
\begin{equation}
  \label{eq:P}
  P_{\mu\nu}(x) = \langle \delta(\eta_{\mu\nu}-x) \rangle \;.
\end{equation}
Figure \ref{fig:density1} shows $P_{0 i}$ for a $8^3\times 4$ lattice
at $\beta=4.5$.  
We adjust $c_{\mu\nu}$ to the minimum of $P_{\mu\nu}$, which is
suppressed by several orders of magnitude with respect to both maxima,
so the precise choice of $c$ has negligible impact, and twist can
indeed be defined unambiguously, as claimed above.  This has been
confirmed numerically by comparing the above choice of $c$ with $c=0$.
The results agree within statistical errors.

The strong suppression of intermediate values of $\eta_{0 i}$ turns out
to be a practical problem: a simple Metropolis algorithm is not capable
of changing the electric twist with sufficient probability on lattices
larger than $4^4$ sites.  To achieve ergodicity across twist sectors,
we use a variant of the multicanonical algorithm \cite{multican},
applied to $\eta_{0 i}$ instead of the action (a ``multitwist''
algorithm so to speak).  The depletion in the density of states as a
function of $\eta_{0 i}$ is removed by adding a bias $\Delta
S(\eta_{01},\eta_{02},\eta_{03})$ to the action, so configurations are
generated with probability $\tfrac1Z\ee^{-S-\Delta S}$.  The
Monte-Carlo process then amounts to a free diffusion across twist
sectors, with dynamics accelerated exponentially.  The modification in
probability is corrected by an inverse factor $\ee^{\Delta S}$ in the
observables.

\begin{figure}[tbp]
  \centering
  \mbox{\includegraphics[keepaspectratio=false,width=12cm,height=90mm]{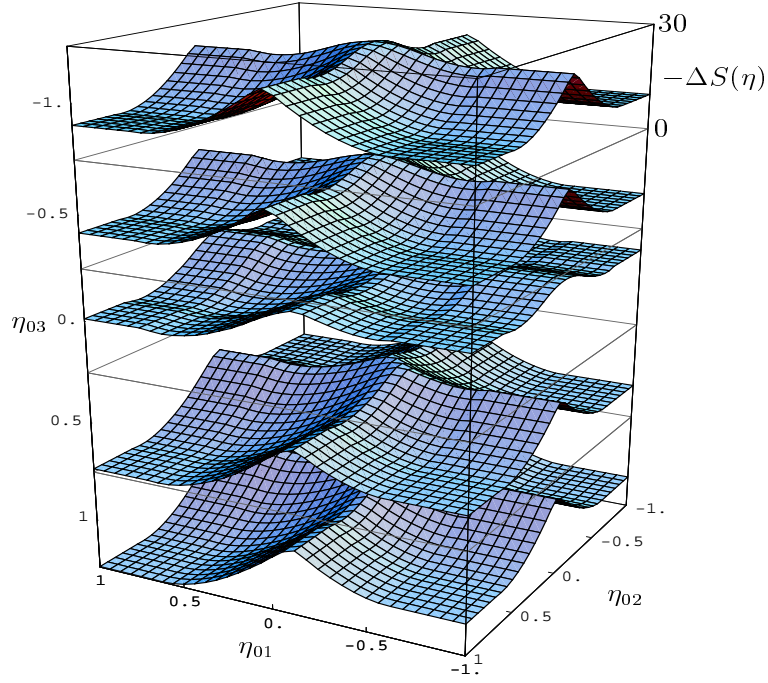}}
  \caption{The three-dimensional reweighting table used in our multicanonical
    Monte Carlo.  Five sections $-\Delta S(\eta_{0 1},\eta_{0 2})$ are
    shown for fixed $\eta_{0 3}$.  No twist is at bottom left, twist in
    all 3 planes at top right.  The table enhances the probability of
    sampling the saddles between twist sectors. Its entries
    $\ee^{-\Delta S}$ vary by 12 orders of magnitude ($8^3\times 4$
    lattice, $\beta=4.5$).}
  \label{fig:bias3d}
\end{figure}

\begin{figure}[tbp]
  \centering
  \mbox{\includegraphics[height=7cm,trim=0 0 0 5]{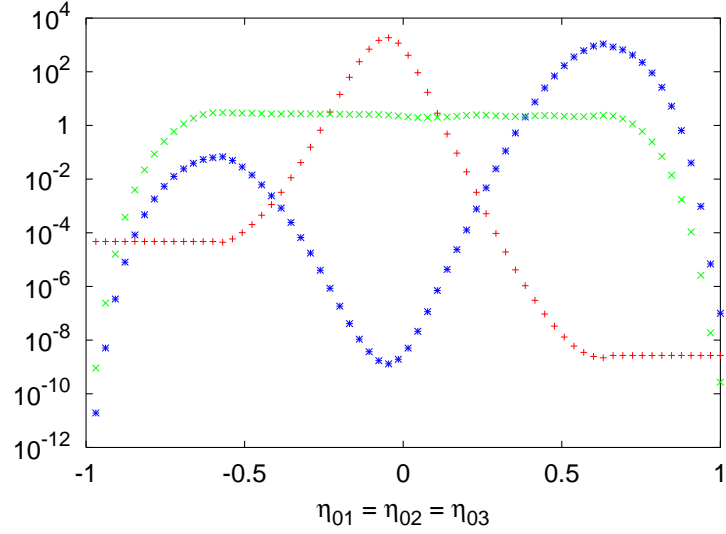}}
  \caption{Cut of the three-dimensional reweighting table along its
    diagonal (red +).  The result of multicanonical Monte Carlo
    sampling is a nearly flat histogram (green x). \protect\nolinebreak
    The density of states shows the dominant twist-0 and the smaller
    twist-3 sectors (blue *).}
  \label{fig:diag}
\end{figure}

The bias $\Delta S(\eta_{01},\eta_{02},\eta_{03})$ can be represented
as a 3-dimensional table since $\eta_{0 i}$ only takes $L^2+1$ discrete
values (cf.~Eq.~(\ref{eq:twist-so3})).  This table is constructed
iteratively, using converged values on small lattices to form starting
values on larger ones.  Since the barrier is expected to grow
approximately quadratically in the system size, the initial bias is
scaled as $\Delta S_{L'}=(L'/L)^2\Delta S_L$.  Such a table is
displayed in Fig.~\ref{fig:bias3d},
for an $8^3 \times 4$ lattice at $\beta=4.5$.  If $\Delta S$ were a sum
of 3 functions of only one variable, the iterative determination would
be greatly simplified.  Alas, Fig.~\ref{fig:bias3d} shows that this is
not the case.  Note that, although $\beta$ has been chosen as small as
possible (the bulk transition to the strongly coupled phase occurs at
$\beta\approx 4.45$), the necessary enhancement of the saddle points
reaches $10^{12}$, as seen more clearly in Fig.~\ref{fig:diag}, 
which is a 1-dimensional cut of the same table along its diagonal (from
$\eta_{0 i}=(-1,-1,-1)$ to $(1,1,1)$).  There, the resulting flatness
of the Monte Carlo sampling probability is clearly visible.  The
measured density of states shows two well-separated peaks,
corresponding to twist sectors $(1,1,1)$ (analogous to
$SU(2)_{\text{pbc}}$) and $(-1,-1,-1)$ (analogous to
$SU(2)_{\text{tbc}}$ in all 3 temporal planes).  The strong suppression
of the saddle point confirms that ordinary Monte Carlo sampling would
remain hopelessly ``stuck'' in one sector, leading to the ``phases''
observed in earlier studies.  However, in spite of the great
multicanonical acceleration, the Monte Carlo evolution of the twist
variables $\eta_{0 i}$ is still slow, and simulating a $10^3 \times 4$
lattice remains beyond the edge of our computer resources.

This prevents us from reaching spatial sizes large enough for a
reliable determination of the dual string tension \cite{tH1}.
Nevertheless, we can still compare the free energies for various small
volumes with those obtained in $\SU(2)$ with the method of
\cite{dFDEP}.  This also allows us to calibrate the $SO(3)$ lattice
spacing.

\begin{figure}[tb]
  \centering
  \enspace\mbox{\includegraphics{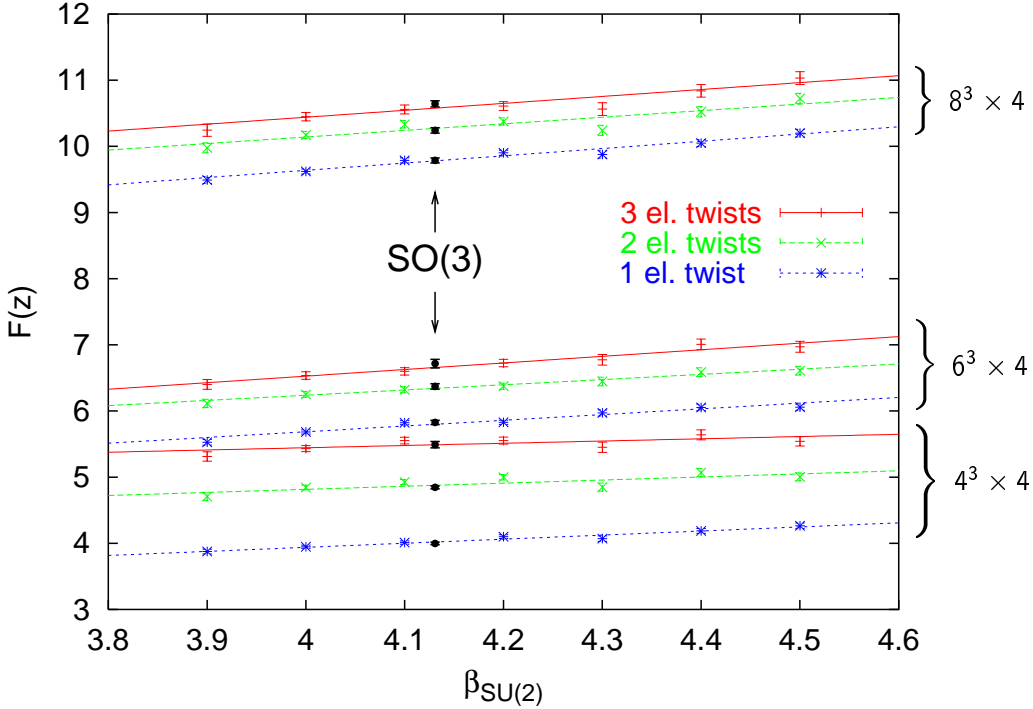}}
  \caption{Comparison of the twist free energies $F(z)$ between 
    $SO(3)$ at $\beta=4.5$ and $SU(2)$ at various $\beta$.  Electric
    twist in 1, 2 and 3 planes is considered, on lattices of size
    $4^3\times4$, $6^3\times4$ and $8^3 \times 4$. One finds
    $\beta_{SO(3)}=4.5 \longleftrightarrow \beta_{SU(2)}=4.13(3)$.}
  \label{fig:match}
\end{figure}

The twisted free energies $F(z)$ are continuum quantities, which depend
upon the spatial size $L$ and the temperature $T$. If the $SO(3)$ and
$SU(2)$ theories represent the same continuum physics, then to each
$\beta_{SO(3)}$ should correspond a value $\beta_{SU(2)}$, which yields
the same lattice spacing and thereby the same $F(z)$ for equal lattice
sizes, modulo small lattice artifacts. We test this statement in
Fig.~\ref{fig:match}.  
The 9 free energies $F(z)$ for 1, 2 or 3 electric twists measured on
$4^3\times4$, $6^3\times4$ and $8^3 \times 4$ lattices in $SO(3)$ at
$\beta_{SO(3)} = 4.5$ are compared with similar quantities measured in
$SU(2)$ at different values of $\beta_{SU(2)}$.  The straight lines
show a linear interpolation in $\beta_{SU(2)}$ of the $SU(2)$ data. One
observes a very good match ($\chi^2/\text{dof}\approx1.35$) of all 9
observables, for $\beta_{SU(2)} \approx 4.13(3)$.

Thus, $a^{-1}(\beta_{SO(3)}=4.5) \approx
a^{-1}(\beta_{SU(2)}=4.13)$, which is about $200$ GeV!  As
conventional wisdom asserts, our $SO(3)$ lattice is very fine indeed.
It cannot be made coarser because of the bulk transition at
$\beta\approx4.45$.  Therefore, to reach low temperatures $T < T_c$ and
probe the confined phase would require a lattice of size
$\mathcal{O}(700^4)$, far beyond what is currently achievable.  We
would like to emphasise that the scale of $200$ GeV is in no way
related to continuum physics.  It is merely due to the bulk phase
transition of the $\SO(3)$ lattice gauge theory beyond which lattice
artifacts dominate, and gives a lower bound on the cutoffs one can use.
This value can be shifted by suppressing (or enhancing) lattice
artifacts \cite{BBMP2}.

Note the closeness of the matched $SO(3)$-$SU(2)$ bare couplings
($\frac{4}{g^2} = 4.5$ vs.\ $4.13$). This should come as no surprise.
Lattice perturbation theory is identical between the $SU(2)$ Wilson
action and the $SO(3)$ Villain action: the difference resides in the
$\alpha$-monopoles, which do not appear in the perturbative
expansion.  Therefore, $\Lambda_{\rm{lattice}}$ is the same in both
theories, and one should expect similar values for the
non-perturbatively matched $\beta$'s. The $\alpha$-monopoles disorder
the $SO(3)$ theory slightly, which is why the $SU(2)$ matching $\beta$
is slightly smaller.

\section{Discussion and conclusions}

\begin{figure}[tb]
  \centering
  \mbox{\includegraphics[height=.45\textwidth]{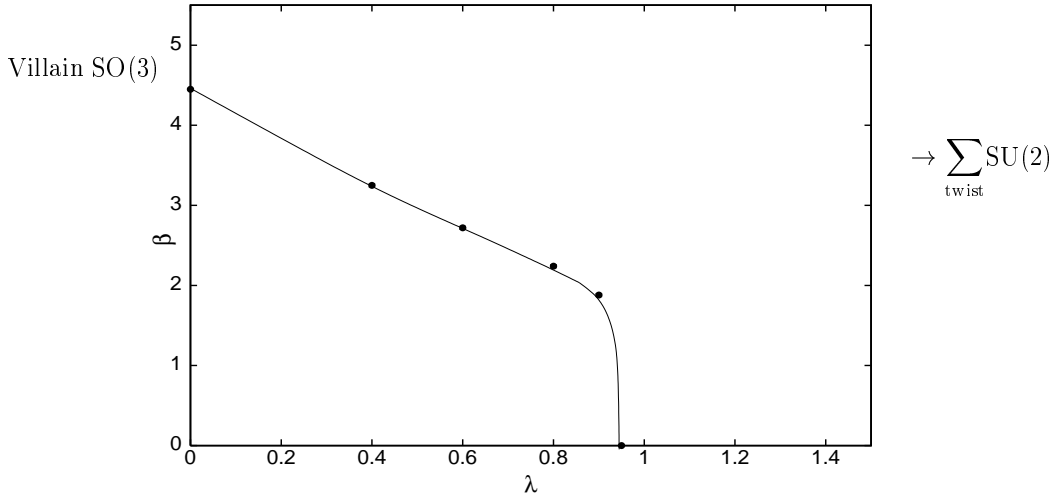}}
  \caption{Phase diagram of Villain SO(3) with $\alpha$-monopole
    suppression, from \cite{DG2}.}
  \label{fig:suppress}
\end{figure}

We have seen that a Monte-Carlo simulation of the confined phase of
SO(3) lattice gauge theory with the Villain action is not feasible in
the near future because a bulk phase transition imposes an upper bound
on the lattice spacing.  This is expected to be true of the Wilson
action as well. Since the bulk phase transition is caused by $\Z_2$
monopoles, the question arises whether the confined phase can be
reached with a modified $\SO(3)$ action suppressing them.  A chemical
potential can move the
transition to stronger coupling and larger lattice spacing.  There are
three kinds of monopoles in the Villain theory, $\alpha_c$, $\eta_c$
[cf.\ Eqs.\ (\ref{eq:alpha-c}) and (\ref{eq:eta-c})] and
$\sigma_c\equiv\alpha_c\eta_c$, and one can introduce chemical
potentials for all:\footnote{The effect of $\lambda_\alpha$ and
  $\lambda_\sigma$ (with $\lambda_\eta=0$) has been studied in
  \cite{Brower}.}
\begin{gather}
  \label{eq:Z-suppress}
  Z(\beta,\lambda_\alpha,\lambda_\eta,\lambda_\sigma) 
  = \sum_{\alpha_p=\pm1} \int \prod_l\dd U_l 
  \exp\bigl[ -S(\beta,\lambda_\alpha,\lambda_\eta,\lambda_\sigma)
  \bigr]
  \\\label{eq:S-chemical}
  S(\beta,\lambda_\alpha,\lambda_\eta,\lambda_\sigma)
  = - \frac{\beta}{2} \sum_p \alpha_p \Tr_\F U_p 
  - \lambda_\alpha \sum_c \alpha_c 
  - \lambda_\eta \sum_c \eta_c 
  - \lambda_\sigma \sum_c \alpha_c\eta_c \;. 
\end{gather}
The parameter $\lambda_\alpha$ (for $\lambda_\eta=\lambda_\sigma=0$)
interpolates between Villain SO(3) and Wilson SU(2)
[cf.~(\ref{eq:SO3-Z2})].  It has been studied by Halliday and Schwimmer
\cite{HS2} who found that the bulk transition is indeed moved to
smaller $\beta$ and disappears ($\beta=0$) around
$\lambda_\alpha^{\text{c}}\approx0.95$, see Fig.~\ref{fig:suppress}.
The deconfinement transition can therefore be studied on small lattices
for $\lambda_\alpha>\lambda_\alpha^{\text{c}}$ \cite{DG2}.  However,
since the limit $\lambda_\alpha\to\infty$ reduces to the SU(2) theory,
it is clear that $\lambda_\alpha\ne0$ reintroduces the $\Z_2$ part of SU(2)
missing in SO(3).  For $\lambda_\alpha\ne0$, (\ref{eq:S-chemical}) is
not really an SO(3) action, but an SU(2) action in disguise.  It is not
suited for verifying the equivalence of SO(3) and SU(2).  This is
expected to be true also of the theory with $\lambda_\sigma\ne0$, since
this also affects the variables $\alpha_p$ such that they cannot be
summed over trivially.

Note that $\lambda_\alpha$ provides an analytic path between the weak
coupling phase of Villain SO(3) and SU(2), since no phase transition
except the above-mentioned monopole condensation transition has been
found \cite{HS2,DG2}.  This is a strong argument in favour of the
equivalence of the two theories in the continuum limit.  It is likely
that an analytic path can be found in the same way for other SO(3)
actions (e.g.\ the Wilson action): any SO(3) theory can be formulated
with auxiliary plaquette variables \`a la Villain, and related to an
SU(2) theory by a chemical potential for $\alpha$-monopoles.  The
action of this SU(2) theory is to some extent arbitrary: only the
centre-symmetric part of the plaquette weight $\exp[-S_p(U_p)]$ is
required to be equal to that of the SO(3) theory.  The anti-symmetric
part is only restricted by the positivity of the weight.  We expect the
monopole condensation transition to move to small $\beta$ with a
suitable choice of antisymmetric part for most theories, so that an
analytic path will exist.

The remaining option, $\lambda_\eta\ne0$ (but
$\lambda_\alpha=\lambda_\sigma=0$) does not reintroduce SU(2) degrees
of freedom: $\alpha_p$ can still be summed over, and one obtains a
manifestly centre-invariant formulation in terms of SO(3) matrices.
$\lambda_\eta$ can also be introduced in the Wilson SO(3) action.  There, like
$\lambda_\alpha$ in the Villain theory, it shifts the bulk transition
to smaller, eventually negative $\beta$ \cite{BBMP1}.  One expects
$\lambda_\eta$ to have the same effect in the Villain theory also.
However, for $\lambda_\eta\to\infty$ the theory now becomes equivalent
to an SU(2) positive plaquette model: if $\eta_c\equiv1$, the plaquette
field $\eta_p$ can be expressed in terms of a link field,
$\eta_p=\prod_{\partial p}\gamma_l$.\footnote{In the presence of twist,
  a representative has to be introduced for each twist, like in
  Eq.~(\ref{eq:Z2-potential-tbc}).}  A change of integration variables
$U_l\to\gamma_l U_l$ makes the plaquette variables trivial,
$\eta_p=\sgn\Tr U_p=+1$.  This is just the positive plaquette
constraint $\Tr U_p\ge0$.  After summing over $\alpha_p$ (still
$\lambda_\alpha=0$), we obtain a positive plaquette model with action
$S=-\sum_p\bigl\{\frac{\beta}{2}\Tr U_p + \ln[1+\exp(-\beta\Tr
U_p)]\bigr\}$.  This modification of action is not expected to have a
significant effect on the deconfinement transition.  Therefore,
comparing SU(2) with Wilson action and $\SO(3)$ with action
$S(\beta,0,\lambda_\eta,0)$, $\lambda_\eta>0$ is little more than
comparing the former with a positive plaquette version of itself.  For
the latter, equivalence with Wilson SU(2) seems established \cite{FHM}.
Moreover, maintaining ergodicity is even more difficult than without
monopole suppression because the barriers between different twist
sectors are even higher.

We conclude that the equivalence of SO(3) and SU(2) lattice gauge
theory cannot be directly verified by simulation at the present time.
The arguments presented in the literature against such an equivalence, 
however, can be
invalidated: an analytic path between the theories at weak coupling
exists under weak assumptions and the phase with negative adjoint
Polyakov loop observed in SO(3) turns out to correspond to sectors with
electric twist in SU(2).  The only difference between the two theories
is that a definite twist sector is selected via the choice of boundary
conditions in $SU(2)$, whereas all sectors are automatically summed
over within periodic boundary conditions in $SO(3)$.  Also the absence
of the centre symmetry in SO(3) is not a problem: it is not needed to
characterise the deconfinement transition.  The (spatial) centre vortex
-- or electric twist -- free energy can be used as an order parameter
in both theories.  At the very high temperatures accessible to
simulations to date, the centre vortex free energies of SO(3) and SU(2)
are numerically compatible, provided the scales are adjusted between
both theories.

Finally, one can try to generalise the lesson learnt here about the
structure of the gauge group $G$ necessary for confinement. $SO(3)$
shows that a non-trivial centre is not required (cf.~\cite{SY}). On the
other hand, the existence of the dual string tension characterising the
deconfined phase arises from that of~'t~Hooft loops or twist sectors.
Those in turn follow from the non-trivial first homotopy group
$\pi_1[G/\Centre(G)]$, which is the same for $SU(2)$ and $SO(3)$.  A
non-trivial $\pi_1[G/\Centre(G)]$, or more precisely $\pi_1[G/Z_G]$,
where $Z_G$ is the discrete part of the centre of $G$, is necessary for
a dual string tension to be defined.  The non-trivial elements of this
group are the centre vortex (or twist) excitations.  There are only 3
(simple) non-Abelian Lie groups for which $\pi_1[G/Z_G]$ is trivial:
$G_2$, $F_4$ and $E_8$.  The absence of a dual string tension in the
corresponding gauge theories prompts speculations about the nature of
the confinement/deconfinement phase transition.  The study of G$_2$
proposed in \cite{HMPW} will be interesting in this connection.

\section{Acknowledgements}
We are grateful to A.~Barresi, G.~Burgio, M.~M\"uller-Preussker and
L.~von Smekal for discussions.

\appendix

\section*{Appendix:\quad Twist and Polyakov loops}
\label{sec:PL}

Here, we shall review the relation between twist and Polyakov loops
for general twist eaters, working out the arguments sketched in
\cite{GA}.  

Consider a twist eater (\ref{eq:twist-eater}) given in terms of 4
constant matrices $\Gamma_\mu\in\SU(2)$ which satisfy the relations
(\ref{eq:t-e-conditions}),
\begin{equation}
  \label{eq:twist-group}
  \Gamma_\mu \Gamma_\nu = \ee^{\ii\pi n_{\mu\nu}} \Gamma_\nu\Gamma_\mu
\end{equation}
with $n_{\mu\nu}=0,1$.  These relations define a group, and the
matrices $\Gamma_\mu$ can be considered a (non-unique) 2-dimensional
representation of this group.  The simple argument below depends on the
fact that, for $\SU(2)$, all such representations are irreducible
unless $n_{\mu\nu}=0$ for all $\mu$, $\nu$.

A generalised Polyakov loop in the fundamental representation which
winds around several directions, successively $s_\mu$ times around the
$\mu$-direction, can be defined as
\begin{align}
  \label{P(s)}
  P_\F(s) &= \tfrac12\Tr_\F \Gamma(s) \\
  \label{Gamma(s)}
  \Gamma(s) &= \Gamma_3^{s_3} \Gamma_2^{s_2} \Gamma_1^{s_1} \Gamma_0^{s_0}
  \;.
\end{align}
It is not necessary to consider other orders of the factors
$\Gamma_\mu$, since, by Eq.~(\ref{eq:twist-group}), the order only
affects the sign of $P_\F(s)$.  Furthermore, irreducibility implies
$\Gamma_\mu^2=\pm1$ \cite{GA}, so $|P_\F(s)|$ depends on $s_\mu$
only modulo~2 and it is sufficient to consider $s_\mu\in\{0,1\}$.

We show that, provided $n\ne0$, $|P_\F(s)|$, and thus
$P_\A(s)$, only depend on $n_{\mu\nu}$ and not on the particular
representation $\Gamma_\mu$.  More precisely,
\begin{equation}
  \label{eq:P-twisteater}
  |P_\F(s)| = 
  \begin{cases}
    0 & \mbox{if } \sum_{\nu} n_{\mu\nu} s_\nu \ne 0 \pmod2 \\
    1 & \mbox{if } \sum_{\nu} n_{\mu\nu} s_\nu = 0 \pmod2 \;. \\
  \end{cases}
\end{equation}

To see this, note that Eq.~(\ref{eq:twist-group}) implies 
\begin{equation}
 \Gamma(s') \Gamma(s) = \ee^{\ii\pi s'\cdot n s} \Gamma(s) \Gamma(s')
\end{equation}
where $s'\cdot n s=\sum_{\mu\nu} s'_\mu n_{\mu\nu} s_\nu$.  Now, $n
s\ne0\pmod2$ implies that $s'\cdot n s=1\pmod2$ for some $s'$ and thus
\begin{equation}
  \label{eq:anti}
  \Gamma(s') \Gamma(s) = - \Gamma(s) \Gamma(s') \;.
\end{equation}
There is a theorem that says that the relation (\ref{eq:anti}) for two
invertible matrices $\Gamma(s)$ and $\Gamma(s')$ implies
$\Tr_\F\Gamma(s)=0$ (Lemma 1 of \cite{GA}, Sec.~4.3).  This proves
the first part of (\ref{eq:P-twisteater}).  If $n s=0\pmod2$, on the
other hand, $\Gamma(s)$ commutes with all $\Gamma(s')$, in particular
the generators $\Gamma_\mu$.  If $n\ne0$, irreducibility implies that
$\Gamma(s)=\pm1$, proving the second part.

It is straightforward to convince oneself that the values of
$|P_\F(s)|$ for $s_\mu=0,1$ determine the twist $n_{\mu\nu}$
uniquely through Eq.~(\ref{eq:P-twisteater}), see Table \ref{tab:PL}.

For vanishing twist, all values of the Polyakov loop have minimal
classical action, since (\ref{eq:twist-group}) is satisfied by
arbitrary diagonal matrices $\Gamma_\mu$.  These are the so-called
``toron'' modes \cite{torons}.

\begin{table}[htb]
  \centering
  \begin{tabular}{cccccccccc}
    $|\vec m|^2$ & $|\vec k|^2$ & \hspace{1em} & 
    $P_0$ & $|\vec P|^2$ & $\vec P\cdot\vec M$ &
    $|\vec M|^2$ & $\vec M\cdot\vec K$ &
    $|\vec K|^2$ & $\vec P\cdot\vec K$ \\\hline
    0 & 1 && 0 & 2 & 0 & 1 & 0 & 0 & 0 \\
    0 & 2 && 0 & 1 & 1 & 1 & 0 & 0 & 0 \\
    0 & 3 && 0 & 0 & 0 & 3 & 0 & 0 & 0 \\\hline
    1 & 0 && 1 & 1 & 0 & 0 & 0 & 1 & 1 \\
    1 & 1 && 0 & 1 & 0 & 0 & 0 & 1 & 0 \\
    1 & 2 && 0 & 1 & 0 & 0 & 0 & 0 & 0 \\\hline
    2 & 0 && 1 & 0 & 0 & 1 & 0 & 0 & 0 \\
    2 & 1 && 0 & 0 & 0 & 1 & 0 & 2 & 0 \\
    2 & 2 && 0 & 0 & 0 & 1 & 1 & 1 & 0 \\
    2 & 3 && 0 & 0 & 0 & 1 & 0 & 0 & 0 \\\hline
    3 & 0 && 1 & 0 & 0 & 0 & 0 & 0 & 0 \\
    3 & 2 && 0 & 0 & 0 & 0 & 0 & 1 & 0
  \end{tabular}
  \caption{Multiple Polyakov loops for twist eaters with twist  
    $n_{i j}=\varepsilon_{i j k} m_k$, $n_{0 i}=k_i$.  Polyakov loops
    winding around up to 2 cycles of the torus are considered:
    $P_\mu=|P_\F(e_\mu)|$, $K_i=|P_\F(e_0+e_i)|$ and
    $M_k=|P_\F(e_i+e_j)|$ with $\{i,j,k\}=\{1,2,3\}$.
    This information is sufficient to determine the twist up to a
    permutation of space coordinates.  The individual components can be
    obtained from the components of $\vec P$, $\vec K$, $\vec M$ and
    $P_0$, if desired.  Some twists are missing because twist eaters
    only exist for orthogonal twist.}
  \label{tab:PL}
\end{table}

\providecommand{\href}[2]{#2}\end{document}